\documentclass[usenatbib]{mn2e}

\usepackage{psfig,morefloats,url}

\footnotesize
\newdimen\digitwidth    
\setbox0=\hbox{\rm0}
\digitwidth=\wd0
\catcode`!=\active
\def!{\kern\digitwidth}
\normalsize

\title[PSR~J1453+1902] {PSR~J1453+1902 and the radio luminosities of
solitary versus binary millisecond pulsars}
\author[D.R.~Lorimer et al.]
{D.R.~Lorimer$^{1}$,\thanks{Email: Duncan.Lorimer@mail.wvu.edu}
M.A. McLaughlin$^{1}$, D.J. Champion$^2$ and I.H.~Stairs$^3$
\\
$^1$Department of Physics, West Virginia University, PO~Box~6315, Morgantown,
WV~26506, USA\\
$^2$McGill University Physics Department, Montreal, QC H3A2T8, Canada\\
$^3$Department of Physics \& Astronomy, University of British Columbia,
6224 Agricultural Road, Vancouver, B.C. V6T 1Z1, Canada
}
\let\leqslant=\leq 
\date{Accepted 2007 May 4. Received 2007 May 3; in original form 2007 April 15}
\begin{document}

\maketitle
\newcommand{\setthebls}{
}

\setthebls

\begin{abstract} 
We present 3~yr of timing observations for PSR~J1453+1902, 
a 5.79-ms pulsar discovered during a 430-MHz drift-scan survey
with the Arecibo telescope. Our observations
show that PSR~J1453+1902 is solitary and has a proper motion
of $8 \pm 2$~mas~yr$^{-1}$. At the nominal distance of 1.2~kpc
estimated from the pulsar's dispersion measure, this corresponds to 
a transverse speed of $46 \pm 11$~km~s$^{-1}$, typical of 
the millisecond pulsar population. We analyse
the current sample of 55 millisecond pulsars in the Galactic
disk and revisit the question of whether the luminosities of isolated
millisecond pulsars are different from their binary counterparts.
We demonstrate that the apparent differences in the luminosity
distributions seen in samples selected from 430-MHz surveys can
be explained by small-number statistics and observational selection
biases. An examination of the sample from 1400-MHz surveys
shows no differences in the distributions. The simplest conclusion
from the current data is that the spin, kinematic, spatial
and luminosity distributions of isolated and binary millisecond
pulsars are consistent with a single homogeneous population.
\end{abstract}


\section{INTRODUCTION}\label{sec:intro}

Twenty-five years after the discovery of the first millisecond pulsar 
\citep{bkh+82}, the sample of these objects currently known is now
close to 200, with the majority being found in searches of
globular clusters \citep[for a review, see][]{cr05}. While searches 
in clusters are far from straightforward, finding millisecond pulsars
in the Galactic disk is a difficult endeavour due to the dispersive
and scattering effects of the interstellar medium which hamper their
detection. Indeed, only 55 out of roughly 1500 pulsars (4\%) currently 
known in the Galactic disk are millisecond pulsars. Despite this low
fraction, the numbers are now at the level where statistically 
significant trends can be identified in the sample and inferences
made about the underlying population.

One such example is the apparent difference in luminosities between
isolated and binary millisecond pulsars, first noted by 
\citet{bjb+97} from 430-MHz observations,
in which isolated millisecond pulsars were
on average fainter than their binary counterparts. This trend was
also seen by \citet{kxl+98} in 1400-MHz data. More recently,
\citet{lkn+06} revisited this issue from a different perspective.
They found that, while the velocity distribution of the
isolated millisecond pulsars is compatible with that of binary
systems, there appears to be a 
difference in the distribution of heights above the Galactic plane for
the two populations, with solitary millisecond pulsars residing closer
to the plane than the binary systems. As discussed by \citet{lkn+06},
given identical velocity dispersions, the only way to explain the
different scale heights would be if the isolated millisecond pulsars
are truly fainter on average and therefore harder to detect further from
the Earth and hence closer to the Galactic plane. If the luminosity
difference is a real effect, then it represents an important clue to
the origin of millisecond pulsars.

To increase the sample of millisecond and binary pulsars in the
Galactic disk, we have been carrying out drift-scan surveys
with the Arecibo telescope \citep{lma+04,clm+05,mlc+05,lxf+05}. Two
isolated millisecond pulsars were found in this survey: PSR~J1944+0907
\citep{clm+05} and PSR~J1453+1902 which was briefly discussed by
\citet{mlc+05}. In this paper, we present the discovery and detailed
follow-up observations of PSR~J1453+1902 and
take the opportunity to compare what is currently known about the
population of solitary and binary millisecond pulsars.  In Section
\ref{sec:disctiming} we describe our observations and present a timing
ephemeris for PSR~J1453+1902. As shown in Section \ref{sec:noplanets},
these timing measurements rule out the presence of any Earth-mass 
companions around this pulsar.  In Section \ref{sec:comparison}, we
carry out a comparison of the solitary and millisecond pulsar
samples. Our conclusions are summarized in Section \ref{sec:conclusions}.

\section{Discovery and timing of PSR~J1453+1902}\label{sec:disctiming}

The 5.79-ms pulsar~J1453+1902 was one of eleven pulsars found during a
430-MHz survey with the Arecibo telescope
\citep{mla+03,lma+04,mlc+05}.  The survey observations were carried
out in drift-scan mode using the Penn State Pulsar Machine (PSPM), a
$128 \times 60$-kHz channel 
analogue spectrometer, to acquire radio signals from the
430-MHz line feed at Arecibo with 4-bit precision every
80$\mu$s. After combination of the two orthogonal circular
polarizations, total-power data were written to tape for off-line
processing. In this mode, a point on the sky drifted through the
10-arcmin primary beam in about 40~s and data were collected along
strips of constant declination $\delta$ at a rate of 60~$\cos
\delta$~deg$^2$ per day. Roughly 1700~deg$^2$ of sky was covered
during the survey.

\begin{figure}
\psfig{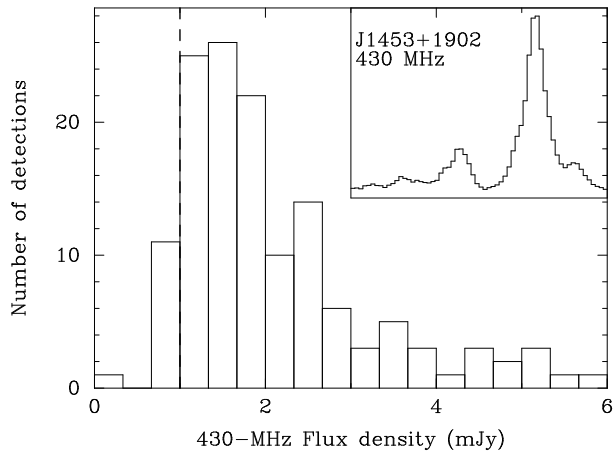}
\caption{
Distribution of 430-MHz flux densities for PSR~J1453+1902.
The dashed line shows the detection threshold for this pulsar
in our original search-mode observation. Inset:
integrated 430-MHz profile showing 360 degrees of
rotational phase. The profile was
produced by phase-aligning and summing all the 430-MHz
detections. The equivalent integration time is 39.3~hr and
effective time resolution (calculated from the quadrature
sum of the 80~$\mu$s sampling time and the dispersive smearing across
a 60-kHz PSPM frequency channel) is 149~$\mu$s.
\label{fig:prof+flux}
}
\end{figure}

The data were subsequently searched for periodic and transient events
off-line using freely-available analysis
tools\footnote{\url{http://sigproc.sourceforge.net}} following the
procedure described by \citet{lma+04}.  PSR~J1453+1902 was originally
detected at a dispersion measure (DM) of 14.2~cm$^{-3}$~pc with a
signal-to-noise ratio of 27 in the amplitude spectrum of the Fourier
transform from data taken on 1998, January
19. Follow-up timing observations were carried out on all the pulsars
from the survey using the 430-MHz Gregorian receiver at Arecibo and
the PSPM as described in detail by \citet{clm+05}.  For
PSR~J1453+1902, all data were taken in search mode and folded off-line
using a preliminary ephemeris derived from the original search
detection. Pulse time-of-arrival (TOA) measurements were obtained from
the folded data by convolving each integrated profile with a high
signal-to-noise template formed from many independent observations
\cite[for further details on the timing procedure, see, e.g.,][]{lk05}.
The final template used in the analysis is shown in
Fig.~\ref{fig:prof+flux}. As seen in some other millisecond pulsars,
\cite[see, e.g.,][]{kxl+98}, the pulse shape is complex with significant
amounts of emission over most of the pulse.

A typical observing session consisted of several 10-min observations
of the pulsar. These profiles were subsequently phase-aligned using an
interim timing model to form a single daily profile.  A phase-coherent
timing observation was finally obtained from 62 TOAs spanning a
three-year period (MJD range 52768--53905) using the \textsc{tempo2}
software
package\footnote{\url{http://www.atnf.csiro.au/research/pulsar/tempo2}}
\citep{hem06}.
This process yields an excellent fit to the TOAs using seven free
parameters: right ascension ($\alpha$), declination ($\delta$), proper
motion in right ascension ($\mu_{\alpha}$), proper motion in
declination ($\mu_{\delta}$), spin period ($P$), period derivative
($\dot{P}$) as well as an arbitrary pulse phase shift. To account for
unknown systematic effects, individual TOA uncertainties were
multiplied by a factor of 1.7 to ensure a reduced $\chi^2$ value of
unity. The resulting model minus computed post-fit weighted residuals
were free from systematic trends with a root-mean-square value of
5.2~$\mu$s. The parameter values and their uncertainties are given in
Table 1. The equivalent position in Galactic longitude and latitude
($l$ and $b$) and the composite proper motion
($\mu=\sqrt{\mu_{\alpha}^2+\mu_{\delta}^2}$) are also
listed. The dispersion measure (DM) was obtained in a subsequent call
to \textsc{tempo2} in which the nominal best-fit parameters from the
previous fit were held constant and TOAs derived from four independent
sub-bands across the 7.68~MHz PSPM band were used in the fit.

For each 10-min observation, we estimated the flux density of
PSR~J1453+1902 by measuring the off-pulse DC level of the dedispersed profile
($D$) before it was subtracted. Following \citet{lcx02}, the resulting
profile was converted into mJy units via the scaling factor
$1000\,T/(DG)$, where $T$ is the system temperature and $G$ is the
receiver gain. In these calculations, we assumed a 55~K receiver
temperature plus an additional contribution of 33~K from the sky
background and $G=10$~K~Jy$^{-1}$ for the Gregorian
system.\footnote{Technical details of the receivers at Arecibo can be
found at \url{http://www.naic.edu/~astro/RXstatus}.}  As can be seen
from the distribution of flux densities shown in
Fig.~\ref{fig:prof+flux}, the flux density varies from the mean value
of 2.2~mJy by a factor of 2--3. This is consistent with the modulation
from diffractive scintillation in the interstellar medium. Given the
distribution of flux densities we observe, PSR~J1453+1902 is above the
approximate 1-mJy survey flux density threshold about 90\% of the time.
Although the pulsar is a relatively weak
object, its high Galactic latitude and large angular offset in the sky
from other millisecond pulsars may make it an important addition
to the millisecond pulsar timing array \cite[see, e.g.,][]{jhv+07}.

\begin{table}
\caption{\label{tab:timing}Observed and derived parameters 
for PSR~J1453+1902}

\begin{tabular}{ll}
\hline
\hline
\noalign{\smallskip}
Parameter & Value \\
\noalign{\smallskip}
\hline
\noalign{\smallskip}
Right ascension, $\alpha$ (h:m:s) (J2000) & 14:53:45.7175(1) \\
Declination, $\delta$ (deg:m:s) (J2000) & 19:02:12.224(3) \\
Proper motion in $\alpha$, $\mu_{\alpha}$ (mas yr$^{-1}$) & 3.2(12) \\
Proper motion in $\delta$, $\mu_{\delta}$ (mas yr$^{-1}$) & --6.8(24)\\
Spin period, $P$ (ms) &  5.7923027349664(4) \\
Epoch of period (MJD) &  53337.0 \\
Period derivative, $\dot{P}$ ($\times 10^{-20}$ s s$^{-1}$) & 1.162(3) \\
Dispersion measure, DM (cm$^{-3}$ pc) & 14.049(4) \\
430-MHz flux density, $S_{430}$ (mJy) & 2.2(1)\\
Galactic longitude, $l$ (deg) & 23.395 \\
Galactic latitude, $b$ (deg)  & 60.812\\
50\% pulse width, $w_{50}$ (ms) & 0.4\\
10\% pulse width, $w_{10}$ (ms) & 1.1\\
Composite proper motion, $\mu$ (mas~yr$^{-1}$) & 8(2)\\
\hline
DM-derived distance, $d$ (kpc) & 1.2 \\
Height above the Galactic plane, $z$ (kpc) & 1.0\\
430-MHz luminosity, $L_{430}$ (mJy kpc$^2$) & 3.2\\
Transverse speed, $v_t$ (km~s$^{-1}$) & 46(11)\\
Kinematic bias to $\dot{P}$, $\dot{P}_{\rm kin}$
 ($\times 10^{-20}$ s s$^{-1}$) & 0.11\\
Characteristic age, $\tau$ (Gyr) & 8.0\\
Magnetic field strength, $B$ ($10^8$ G) & 2.5\\
Spin-down luminosity, $\dot{E}$ ($10^{33}$~ergs/s) & 2.1\\
\noalign{\smallskip}
\hline
\end{tabular}
The numbers in parentheses are twice the
quoted uncertainties from \textsc{tempo2} and represent approximately
1-$\sigma$ uncertainties in the measured parameters. Note that \textsc{tempo2}
was invoked with the \verb+-tempo1+ option so that the quoted timing 
parameters are in barycentric dynamical time units.
\end{table}

Also listed in Table~1 are several derived parameters: the distance to
the pulsar in kpc ($d$) estimated from the DM using the Galactic
electron density model of \citet{cl02}, the height above the Galactic
plane ($z=d \sin b$) and 430-MHz luminosity $L_{430} = S_{430} d^2$.
Using the distance estimate and proper motion measurement, we infer
the transverse speed $v_t = \mu d = 46 \pm 11$~km~s$^{-1}$, typical
of the millisecond pulsar population \cite[see, e.g.,][]{hllk05}. The 
proper motion and distance allow us to calculate the kinematic
contribution to the observed $\dot{P}$ \citep{shk70} via
$\dot{P}_{\rm kin} = \mu^2 d P/c$. This amounts to a 10\% overestimate
of the true $\dot{P}$ value and we take this into account when
calculating the characteristic age ($\tau=P/2\dot{P}$), inferred
surface dipole magnetic field strength ($B=3.2\times 10^{19} \sqrt{P
\dot{P}}$~Gauss) and spin-down luminosity ($\dot{E}= 3.95 \times
10^{46} \dot{P}/P^3$~ergs~s$^{-1}$). For further details of the
definitions and assumptions in these parameters, see, e.g., \citet{lk05}.

\section{A search for orbiting companions}\label{sec:noplanets}

\begin{figure}
\psfig{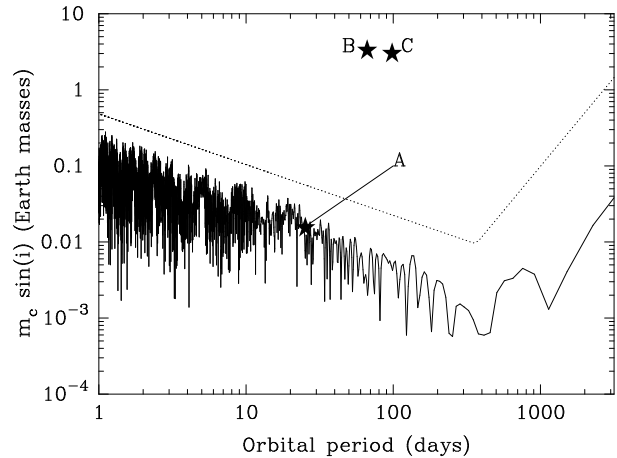}
\caption{
Lomb-Scargle spectrum for the timing residuals of PSR~J1453+1902 shown
as a function of companion mass versus orbital period.  For details of
the calculation, see \citet{fck+03}.  The dotted line shows the 99.9\%
confidence level of the analysis. All features in the spectrum are
well below this limit, indicating that there are no significant
periodicities in the data.\label{fig:lomb} For comparison, the starred
points show the three planets (A, B and C) in the PSR~B1257+12 system.}
\end{figure}

An interesting issue ever since the discovery of planets around
PSR~B1257+12 \citep{wf92} is the lack of any other millisecond pulsar
planetary system in the Galactic disk. The rarity of such planets
makes it important to place limits on newly found millisecond pulsars
by searching for periodic signals in the timing residuals.  Following
the procedure described in detail by \citet{fck+03}, we have carried
out a Lomb-Scargle periodogram analysis \citep{pftv86} on the timing
residuals for PSR~J1453+1902. Although we found no significant 
periodicities in the data, ruling out all Earth-mass planets,
as shown in Fig.~\ref{fig:lomb} we cannot presently exclude smaller
bodies, such as the 0.02~M$_{\earth}$ planet A in the B1257+12 system.

We have carried out a similar analysis on the timing residuals
from the other solitary 5.2-ms pulsar from this survey,
PSR~J1944+0907 \citep{clm+05}. These data are of similar quality
and also reveal no significant signals. Planetary systems 
around pulsars remain a very rare phenomenon.

\section{Millisecond pulsar statistics}\label{sec:comparison}

PSR~J1453+1902 brings the total number of solitary millisecond pulsars
in the Galactic disk to 17. Given the larger sample size available to
us than in the past \citep{bjb+97,kxl+98}, it is appropriate to
revisit the question of whether there is a difference between the 
luminosities of isolated and binary millisecond pulsars. As mentioned in
Section~\ref{sec:intro}, \citet{lkn+06} have recently compared the
velocity distributions of 9 solitary millisecond pulsars and 20
binary millisecond pulsars and find them to be 
consistent.  Surprisingly, however, they also find evidence for a
smaller $z$ height distribution in the isolated pulsar
population. Given the identical velocity distributions, this could
be explained by a difference in luminosity between the two
populations.

We have carried out an updated census of the binary and solitary
millisecond pulsar populations with a view to assessing the
significance of the luminosity distribution difference. We 
therefore compiled a catalogue of flux density and distance
measurements for all currently known millisecond pulsars in the
Galactic disk with $P<10$~ms.
We excluded the millisecond pulsar planetary system
PSR~B1257+12 from this sample, as it may have a separate origin
\citep{brsa93}. The data are summarized in Tables~\ref{tb:binary} and
\ref{tb:isolated}.

Using the Kolmogorov-Smirnoff (KS) statistical test \cite[see,
e.g.,][]{pftv86}, we have compared the samples of isolated and binary
millisecond pulsars in terms of their period, luminosity, spectral
index, transverse speed and spin-down luminosity. The only apparently
statistically significant difference that we found between the samples
was in the luminosity distributions.  We now discuss this issue in
detail.

\begin{figure}
\psfig{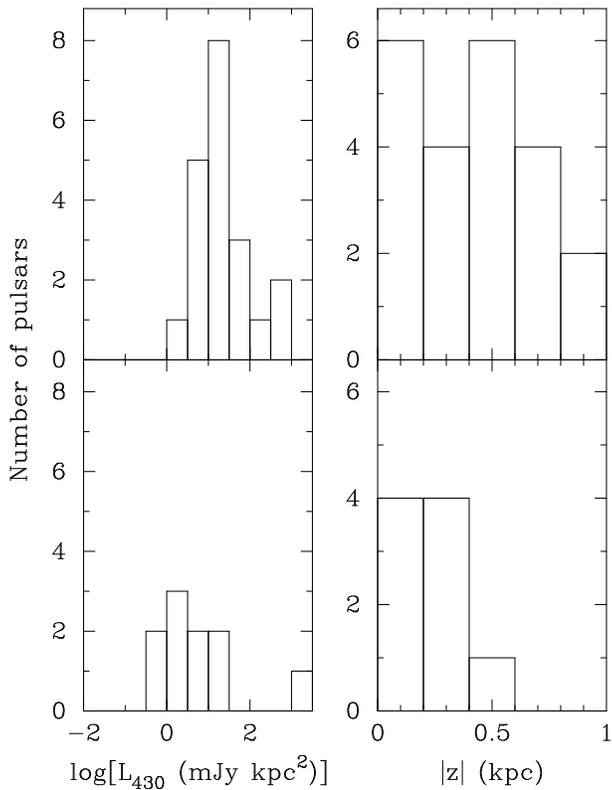}
\caption{
A comparison of the luminosity (left) and $|z|$ distributions
(right) for the sample of 32 Galactic disk millisecond pulsars detected
in 430-MHz surveys. The top panels show the distributions
for the binary millisecond pulsars, while the lower panels
show the distributions for the solitary millisecond pulsars.
The samples appear to be statistically distinct from
one another. \label{fig:l430}
}
\end{figure}

Most of the sky has now been searched for millisecond pulsars at both
430~MHz and 1400~MHz. Given that surveys carried out at these two
frequencies probe different volumes of the Galaxy, we first wish to
assess whether the difference in luminosities is present in the
samples of pulsars detected at each frequency. Fig.~\ref{fig:l430}
shows the luminosity and $z$ distributions for the sample of 32 Galactic disk
pulsars detected by 430-MHz surveys (10 isolated pulsars versus 22
binary pulsars). From a KS test, we see that the luminosity
distributions are different at an apparently statistically significant
level of 99.1\%.  The $z$ distributions appear to be only marginally
different, with a KS test returning an 81.3\% significance level.

\begin{figure}
\psfig{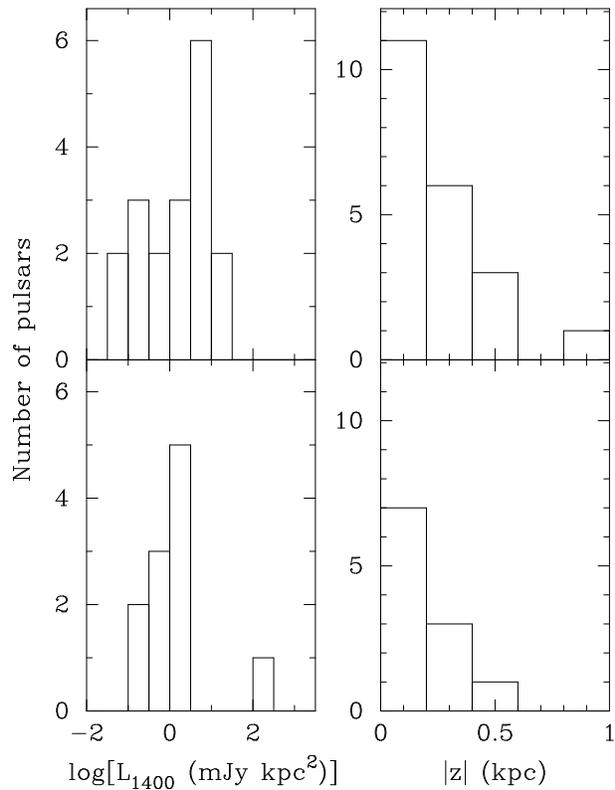}
\caption{
A comparison of the luminosity (left) and $z$ distributions
(right) for the sample of 33 millisecond pulsars detected
in 1400-MHz surveys. The top panels show the distributions
for the binary millisecond pulsars, while the lower panels
show the distributions for the solitary millisecond pulsars.
The samples are statistically indistinguishable.
\label{fig:l1400}
}
\end{figure}

A different conclusion is reached, however, from Fig.~\ref{fig:l1400}
which shows the luminosity and $|z|$ distributions for the sample of
33 pulsars detected by 1400-MHz surveys (11 isolated pulsars versus 22
binary pulsars). To the eye, the distributions appear
consistent. Indeed, the KS tests also show no statistically
significant differences, with the confidence levels for $L$ and $z$
being only 70.9\% and 21.1\% respectively.

There are two possible explanations as to why the luminosity
difference is not seen in both the 430-MHz and 1400-MHz samples.  The
first possibility is that the high-frequency sample does not probe the
luminosity function as deeply as the low-frequency sample.  To
investigate this, we note that the minimum 1400-MHz luminosity in our
sample is 0.1~mJy~kpc$^2$. For a median spectral index of --1.8, the
equivalent 430-MHz luminosity is 1~mJy~kpc$^2$. As this limit is
slightly higher than the observed 430-MHz distribution in
Fig.~\ref{fig:l430}, it remains a tantalyzing possibility that the
effect is only seen in the 430-MHz sample which is more slightly more
sensitive to the low end of the luminosity function than at 1400~MHz.

A second possibility is that the difference is due to a selection
effect.  It is well established \citep[see, e.g.,][]{lml+98} that
430-MHz surveys probe only the local population of millisecond pulsars
out to a distance of 2--3~kpc at most due to propagation effects in
the interstellar medium. As a result, samples of pulsars from these
surveys tend to be stacked in favour of nearby low-luminosity
objects. For any reasonable luminosity function, the high-luminosity
pulsars are rarer objects. If isolated millisecond pulsars are simply
less numerous than their binary counterparts, small-number statistics
will therefore bias the sample in favour of low luminosity objects as
there is a greater chance of having a low-luminosity pulsar in the
sample compared to a higher luminosity one.

To demonstrate this effect, we have carried out a simple simulation of
the millisecond pulsar population using the freely-available
\textsc{psrpop} software package
\citep{lfl+06}\footnote{\url{http://psrpop.sourceforge.net}}. We
generated a model galaxy where millisecond pulsars were distributed
using the radial distribution proposed by \citet{yk04}, an exponential
$z$ distribution with a scale height of 500~pc, a 430-MHz luminosity
distribution with a slope $d \log N / d \log L=-1$ over the range
0.1--100~mJy~kpc$^2$, the period distribution proposed by \citet{cc97}
and intrinsic pulse shapes were approximated as top-hat functions with a
duty cycle of 15\%.  The synthetic pulsars were
then `detected' in the manner as described by \citet{lfl+06} using a
model all-sky 430-MHz survey with similar sensitivity to the Parkes
southern sky survey \cite{lml+98}. To mimic the statistics shown in
Fig.~\ref{fig:l430}, we compared the luminosities of a sample of 10
pulsars randomly selected from our fake all-sky survey and compared
this to another larger sample of 22 objects.

\begin{figure}
\psfig{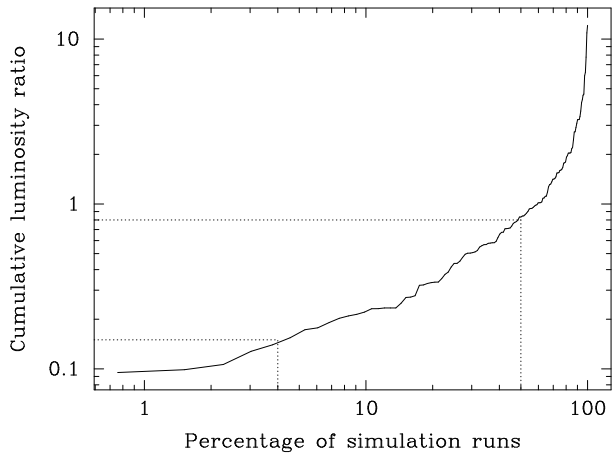}
\caption{
Results of the simulations of samples of millisecond pulsars
detectable by a model 430-MHz all-sky survey. For each simulation
run, we have computed the ratio of the median luminosities for the
10 object versus 22 object samples and display these as a cumulative
distribution plotted against the percentage of all simulation runs.
The observed ratio of median luminosities is 0.15 and is shown by
the lower dotted line. Such small values were produced in about 4\% of
all simulation runs. The upper dotted line shows that, in 50\% of
all cases, the luminosity ratio is less than 0.8.
\label{fig:fake}}
\end{figure}

Due to the nature of random sampling for these relatively small
numbers of objects, we found a considerable variation in the 
results which are shown as a cumulative distribution in Fig.~\ref{fig:fake}.
When averaged over many simulations, the median 430-MHz
luminosity of the sample with 10 pulsars was 20\% lower than the
larger sample, suggesting that this is a significant effect.
The observed ratio of the sample medians is $3.4/22.5=0.15$.
As can be inferred from Fig.~\ref{fig:fake}, such small values
were produced in about 4\% of all simulations.

This simple analysis demonstrates the pitfalls of observational
selection and small-number statistics. It appears that the
probability of observing apparently different luminosity 
distributions from identical parent populations is not as
small as the KS test would suggest. We note that a recent study of the
millisecond pulsar population in globular clusters independently
points to similar luminosity functions for isolated and
binary millisecond pulsars \citep{hrs+07}.

\section{Conclusions}
\label{sec:conclusions}

We have found the isolated 5.79-ms pulsar J1453+1902 during a 
430-MHz drift-scan survey with the Arecibo telescope and measured
its spin and astrometric properties in a dedicated timing campaign.
This completes the timing of the 11 pulsars discovered in this survey
described by \citet{lma+04} and \citet{clm+05}. PSR~J1453+1902
appears to be a typical member of the millisecond pulsar population
and brings the number of isolated pulsars with periods less than 10~ms
currently known in the Galactic disk to 17. 

We have revisited the luminosities of isolated and binary millisecond
pulsars in the Galactic disk
using an up-to-date catalogue of 55 objects. While we confirm
an apparently significant difference in the sample of pulsars detected
by low-frequency surveys, the effect is not present at all in the 
sample detected by 1400~MHz surveys. We demonstrate that this 
could be a selection effect due to isolated millisecond pulsars being 
intrinsically rare by comparison with the binary millisecond pulsars
rather than being intrinsically less luminous. 

Based on the currently available data, we suggest that
isolated and binary millisecond pulsars have consistent spatial,
kinematic and luminosity distributions and there is no longer a need
to posit different origins for the two populations. While we can not
rule out different origins based on this study alone, we can say,
however, that the present data do not {\it require} the populations of
binary and millisecond pulsars to be distinct.

\section*{Acknowledgments} 

The Arecibo observatory, a facility of the National Astronomy and
Ionosphere Centre, is operated by Cornell University in a co-operative
agreement with the National Science Foundation. We thank Alex
Wolszczan for making the PSPM freely available for use at
Arecibo, Michael Kramer for kindly
providing the code used in the Lomb-Scargle analysis in Section 3,
Andrea Lommen for useful discussions and Matthew Bailes for providing
flux density information on PSR~J1721$-$2457.

\begin{table*} 
\caption{The sample of 
38 binary millisecond pulsars with $P<10$~ms currently known in the 
Galactic disk. From left to right, the 
columns list pulsar name, spin period ($P$), binary period ($P_{\rm b}$),
whether the pulsar was detected in 430-MHz and/or 1400-MHz surveys, the
distance ($d$), height above the Galactic plane ($z$),
flux density at 430~MHz ($S_{430}$), 
flux density at 1400~MHz ($S_{1400}$), spectral index (SI),
luminosity  at 430~MHz ($L_{430}$),luminosity  at 1400~MHz ($L_{1400}$)
and references to the literature for the quoted distance and flux measurements.
Those parameters which have not been measured are denoted by the * symbol.
}
\label{tb:binary}
\begin{tabular}{lrrccrrrrrrrrl}
\hline
PSR & \multicolumn{1}{c}{$P$} & \multicolumn{1}{c}{$P_{\rm b}$}
 & \multicolumn{2}{c}{Detected at} &
\multicolumn{1}{c}{$d$} & \multicolumn{1}{c}{$z$}
 & $S_{430}$ & $S_{1400}$ & SI &
$L_{430}$ & $L_{1400}$ & Refs. \\
    & (ms)& (days) & 430 MHz & 1400 MHz & (kpc) & (kpc) & (mJy) & (mJy) & &
\multicolumn{2}{c}{(mJy~kpc$^2$)}& \\
\hline
J0034$-$0534&1.877& 1.6& yes&no&0.54&$-0.50$&17&0.61&$-2.8$&5.0&0.2&1, 2, 2\\
J0218$+$4232&2.323& 2.0& no&no&2.67&$-0.80$&35&0.9&$-3.1$&249.5&6.4&1, 3, 4\\
J0437$-$4715&5.757& 5.7& yes&yes&0.159&$-0.11$&550&142&$-0.9$&13.9&3.6&5, 2, 4\\
J0610$-$2100&3.861& 0.3& no&yes&3.54&$-1.10$&*&0.4&*&*&5.0&1, 6\\
J0613$-$0200&3.062& 1.2& yes&no&0.48&$-0.08$&21&1.4&$-2.3$&4.8&0.3&5, 7, 4\\
\\
J0751$+$1807&3.479& 0.3& yes&no&1.15&$0.41$&10&3.2&$-1.0$&13.2&4.2&1, 8, 4\\
J1012$+$5307&5.256& 0.6& yes&no&0.84&$0.65$&30&3&$-1.9$&21.2&2.1&9, 10, 4\\
J1045$-$4509&7.474& 4.1& yes&yes&1.96&$0.42$&15&3&$-1.4$&57.6&11.5&1, 2, 4\\
J1125$-$6014&2.630& 8.8& no&yes&1.50&$0.02$&*&0.05&*&*&0.1&1, 11\\
J1216$-$6410&3.539& 4.0& no&yes&1.33&$-0.04$&*&0.05&*&*&0.1&1, 11\\
\\
J1435$-$6100&9.348& 1.4& no&yes&2.16&$-0.02$&*&0.25&*&*&1.2&1, 12\\
J1455$-$3330&7.987& 76.2& yes&no&0.53&$0.20$&9&1.2&$-1.7$&2.5&0.3&1, 2, 2\\
J1600$-$3053&3.598& 14.3& no&yes&1.63&$0.46$&*&3.2&*&*&8.5&1, 13\\
J1640$+$2224&3.163& 175.5& yes&no&1.16&$0.72$&*&2&*&*&2.7&1, 4\\
J1643$-$1224&4.622& 147.0& yes&yes&2.41&$0.87$&75&4.8&$-2.3$&435.6&27.9&1, 7, 4\\
\\
J1709$+$2313&4.631& 22.7& yes&no&1.41&$0.75$&2.52&0.2&$-2.1$&5.0&0.4&1, 14, 14\\
J1713$+$0747&4.570& 67.8& yes&yes&0.91&$0.39$&36&8&$-1.2$&29.8&6.6&5, 15, 4\\
J1732$-$5049&5.313& 5.3& no&yes&1.41&$-0.23$&*&*&*&*&*&1\\
J1738$+$0333&5.850& 0.4& no&yes&1.43&$0.44$&*&*&*&*&*&1\\
J1741$+$1351&3.747& 16.3& no&yes&0.92&$0.34$&*&0.93&*&*&0.8&1, 13\\
\\
J1751$-$2857&3.915& 110.7& no&yes&1.10&$-0.02$&*&0.06&*&*&0.1&1, 16\\
J1757$-$5322&8.870& 0.5& no&yes&0.96&$-0.23$&*&*&*&*&*&1\\
J1804$-$2717&9.343& 11.1& yes&yes&0.78&$-0.04$&15&0.4&$-3.1$&9.1&0.2&1, 2, 4\\
J1853$+$1303&4.092& 115.7& no&yes&2.09&$0.20$&*&0.4&*&*&1.7&1, 16\\
B1855$+$09&5.362& 12.3& yes&yes&1.17&$0.06$&31&5&$-1.5$&42.4&6.8&1, 17, 4\\
\\
J1909$-$3744&2.947& 1.5& no&yes&1.14&$-0.38$&*&3.0&*&*&3.9&5, 18\\
J1910$+$1256&4.984& 58.5& no&yes&2.33&$0.07$&*&0.5&*&*&2.7&1, 16\\
J1911$-$1114&3.626& 2.7& yes&no&1.22&$-0.20$&31&0.5&$-3.5$&46.1&0.7&1, 19, 4\\
J1918$-$0642&7.646& 10.9& no&yes&1.24&$-0.20$&*&*&*&*&*&1\\
J1933$-$6211&3.354& 12.8& no&yes&0.52&$-0.25$&*&2.3&*&*&0.6&1, 13\\
\\
B1953$+$29&6.133& 117.3& yes&no&4.64&$0.04$&15&1.1&$-2.2$&322.9&23.7&1, 20, 4\\
B1957$+$20&1.607& 0.4& yes&no&2.49&$-0.20$&20&0.4&$-3.3$&124.0&2.5&1, 21, 4\\
J2019$+$2425&3.935& 76.5& yes&no&1.49&$-0.17$&2.7&*&*&6.0&*&1, 22\\
J2033$+$17&5.949& 56.3& yes&no&2.00&$-0.45$&*&*&*&*&*&1\\
J2051$-$0827&4.509& 0.1& yes&no&1.04&$-0.53$&22&2.8&$-1.7$&23.8&3.0&1, 23, 4\\
\\
J2129$-$5721&3.726& 6.6& yes&no&1.36&$-0.94$&14&1.4&$-1.9$&25.9&2.6&1, 2, 4\\
J2229$+$2643&2.978& 93.0& yes&no&1.45&$-0.64$&13&0.9&$-2.3$&27.3&1.9&1, 15, 4\\
J2317$+$1439&3.445& 2.5& yes&no&0.83&$-0.56$&19&4&$-1.2$&13.1&2.8&1, 24, 4\\
\hline
\end{tabular}

The references used in this compilation are
1: \cite{cl02},
2: \cite{tbms98},
3: \cite{nbf+95},
4: \cite{kxl+98},
5: \cite{hbo06},
6: \cite{bjd+06},
7: \cite{lnl+95},
8: \cite{lzc95},
9: \cite{lcw+01},
10: \cite{nll+95},
11: \cite{lfl+06},
12: \cite{mlc+01},
13: \cite{jbo+07},
14: \cite{lwf+04},
15: \cite{cam95a},
16: \cite{sfl+05},
17: \cite{ffb91},
18: \cite{jbv+03},
19: \cite{llb+96},
20: \cite{bbf+84},
21: \cite{fbb+90},
22: \cite{ntf93},
23: \cite{sbl+96},
24: \cite{cnt96}.
\end{table*}

\begin{table*} 
\caption{The sample of 
17 solitary millisecond pulsars with $P<10$~ms currently known in the 
Galactic disk. From left to right, the 
columns list pulsar name, spin period ($P$), 
whether the pulsar was detected in 430-MHz and/or 1400-MHz surveys, the
distance ($d$), height above the Galactic plane ($z$),
flux density at 430~MHz ($S_{430}$), 
flux density at 1400~MHz ($S_{1400}$), spectral index (SI), 
luminosity  at 430~MHz ($L_{430}$),luminosity  at 1400~MHz ($L_{1400}$)
and references to the literature for the quoted distance and flux measurements.
Those parameters which have not been measured are denoted by the * symbol.
}
\label{tb:isolated}
\begin{tabular}{lrccrrrrrrrrl}
\hline
PSR & \multicolumn{1}{c}{$P$} & \multicolumn{2}{c}{Detected at} &
\multicolumn{1}{c}{$d$} & \multicolumn{1}{c}{$z$}
 & $S_{430}$ & $S_{1400}$ & SI &
$L_{430}$ & $L_{1400}$ & Refs. \\
    & (ms)& 430 MHz & 1400 MHz & & (kpc) & (kpc) & (mJy) & (mJy) & &
\multicolumn{2}{c}{(mJy~kpc$^2$)}& \\
\hline
J0030$+$0451&4.865&  yes&no&0.30&$-0.25$&7.9&0.6&$-2.2$&0.7&0.1&1, 2, 2\\
J0711$-$6830&5.491&  yes&yes&0.86&$-0.34$&10&1.6&$-1.6$&7.4&1.2&3, 4, 5\\
J1024$-$0719&5.162&  yes&yes&0.52&$0.34$&4.6&0.66&$-1.6$&1.2&0.2&6, 4, 5\\
J1453$+$1902&5.793&  yes&no&1.15&$1.00$&2.2&*&*&2.9&*&3, 7\\
J1629$-$6902&6.001&  no&yes&0.96&$-0.23$&*&2.7&*&*&2.5&3, 8\\
\\
J1721$-$2457&3.497&  no&yes&1.29&$0.15$&*&1.8&*&*&3.0&3, 9\\
J1730$-$2304&8.123&  yes&yes&0.53&$0.06$&43&4&$-1.9$&12.1&1.1&3, 10, 5\\
J1744$-$1134&4.075&  yes&yes&0.47&$0.07$&18&3&$-1.5$&4.0&0.7&6, 11, 5\\
J1801$-$1417&3.625&  no&yes&1.52&$0.11$&*&0.17&*&*&0.4&3, 12\\
J1843$-$1113&1.846&  no&yes&1.69&$-0.10$&*&0.10&*&*&0.3&3, 13\\
\\
J1905$+$0400&3.784&  no&yes&1.71&$-0.04$&*&0.050&*&*&0.1&3, 13\\
J1911$+$1347&4.626&  no&yes&2.07&$0.07$&*&0.08&*&*&0.3&3, 12\\
B1937$+$21&1.558&  yes&yes&3.57&$-0.02$&240&10&$-2.7$&3058.8&127.4&3, 14, 5\\
J1944$+$0907&5.185&  yes&no&1.79&$-0.23$&3.9&*&*&12.5&*&3, 15\\
J2010$-$1323&5.223&  no&yes&1.02&$-0.41$&*&1.6&*&*&1.7&3, 16\\
\\
J2124$-$3358&4.931&  yes&no&0.25&$-0.18$&17&1.6&$-2.0$&1.1&0.1&6, 11, 5\\
J2322$+$2057&4.808&  yes&no&0.80&$-0.48$&0.5&*&*&0.3&*&3, 17\\
\hline
\end{tabular}

The references used in this compilation are
1: \cite{lkn+06},
2: \cite{lzb+00},
3: \cite{cl02},
4: \cite{bjb+97},
5: \cite{kxl+98},
6: \cite{hbo06},
7: this paper,
8: \cite{obhv04},
9: Bailes (2007) private communication,
10: \cite{lnl+95},
11: \cite{tbms98},
12: \cite{lfl+06},
13: \cite{hfs+04},
14: \cite{ffb91},
15: \cite{clm+05},
16: \cite{jbo+07},
17: \cite{ntf93}.
\end{table*}
\end{document}